\UseRawInputEncoding
\documentclass[a4paper,11pt]{article}
\pdfoutput=1 

\usepackage{jinstpub} 

\title{\boldmath KM3NeT/ARCA expectations in view of a novel multimessenger study of starburst galaxies}

\author[a,1]{Antonio Marinelli,\note{Corresponding author.}}
\author[a,b]{Antonio Ambrosone,}
\author[a,c]{Walid Idrissi Ibnsalih,}
\author[a,b,e]{Gennaro Miele,}
\author[a]{Pasquale Migliozzi,}
\author[a,b]{Ofelia Pisanti,}
\author[d]{ and Ankur Sharma,}

\affiliation[a]{INFN - Sezione di Napoli, Complesso Univ. Monte S. Angelo, I-80126 Napoli, Italy}
\affiliation[b]{Dipartimento di Fisica ``Ettore Pancini'', Università degli studi di Napoli ``Federico II'', Complesso Univ. Monte S. Angelo, I-80126 Napoli, Italy}
\affiliation[c]{Università degli Studi della Campania "Luigi Vanvitelli", Dipartimento di Matematica e Fisica, viale Lincoln 5,
	Caserta, 81100 Italy}
\affiliation[d]{Department of Physics and Astronomy, Uppsala University
	, Box 516, SE-75120 Uppsala, 
	Sweden}
\affiliation[e]{Scuola Superiore Meridionale, Università degli studi di Napoli ``Federico II'', Largo San Marcellino 10, 80138 Napoli, Italy}

\emailAdd{antonio.marinelli@na.infn.it}

\abstract{Starburst galaxies (SBGs) and more in general starforming galaxies represent a class of galaxies with a high star formation rate (up to 100 $M_{\odot}$/year). Despite their low luminosity, they can be considered as guaranteed ``factories'' of high energy neutrinos, being ``reservoirs'' of accelerated cosmic rays and hosting a high density target gas in the central region. In this contribution we present a novel multimessenger study of these sources and the possibility of observing their neutrino signals with the KM3NeT/ARCA telescope. The differential sensitivity for different SBG scenarios is reported considering track-like neutrino events in the 100 GeV-100 PeV energy range.}


\collaboration[c]{on behalf of the KM3NeT collaboration}

\proceeding{VLVnT2021 -  Very Large Volume Neutrino Telescope Workshop\\
  18 -21 May 2021\\
  Valencia}

\begin{document}
\maketitle
\flushbottom

\section{Introduction}
\label{sec:intro}

Star forming and Starburst Galaxies  (SFGs and SBGs) are astrophysical sources which are characterised by a star formation rate which ranges from 10 to 100 times the one measured in normal galaxies like the Milky way \cite{2006ApJ...645..186T}. The intense star-forming activity implies that a high density of interstellar gas is present in the source core which represents the target for inelastic collision of high-energy protons accelerated by supernova remnants. What is more, these sources are considered to be cosmic-ray reservoirs,  the accelerated particles are likely to be confined within the astrophysical environment and therefore  they have an enhanced probability of emitting gamma rays and neutrinos via hadronic collisions. Different analyses have been carried out about this source class (for instance see \cite{Murase:2013rfa,Tamborra:2014xia}) and demonstrated that it cannot be the sole contribution for IceCube's observations. Despite all the phenomenological constraints, SFGs and SBGs could still play an important role in high-energy neutrino production and explain a seizable part of these observations. In fact, recent works (\cite{Peretti:2019vsj,Ambrosone:2020evo}) have renewed the analysis about this class. In particular, \cite{Peretti:2019vsj} has shown how their flux could be consistent with the one of through-going muon neutrinos, while \cite{Ambrosone:2020evo} has performed a multi-component fit demonstrating that SBGs could explain up to $ 40\%$ of the high energy starting events (HESE) and their contribution is expected to peak at hundred of TeVs. 
The major obstacle encountered when obtaining their ``calorimetric'' parameters it is their low luminosity in the gamma-ray range. In fact, only a dozen of SBGs have been observed as point-like sources of gamma-rays by Fermi-LAT. However, this does not mean their diffuse contribution is negligible with respect to other sources when we consider their emission up to high redshifts as we highlight in this contribution. On the other hand, the IceCube collaboration (\cite{IceCube:2019cia}) has reported a $2.9\sigma$ excess of neutrino events coming from NGC 1068, which is one of the SBG observed by the Fermi-LAT telescope. Moreover \cite{Ambrosone:2021aaw} has recently shown how the cores of SMC and Circinus galaxy could possibily be observed in six years of data taking by KM3NeT/ARCA, which would demonstrate that the star-forming activity is dominated by hadronic emissions. For this reason, the KM3NeT/ARCA telescope could be crucial in order to detect such sources both as point-like objects and  to constrain their diffuse flux. In this context, we present a novel sensitivity study for the possible diffuse signal of this class of sources provided by \cite{Ambrosone:2020evo} considering the phase 2.0 of the incoming KM3NeT/ARCA telescope.
In this study, differentiating the energy range between 100 GeV and 10 PeV into 11 bins and only taking into account up-going track-like events, we calculate the KM3NeT/ARCA differential sensitivity for the expected diffuse SBG emission up to a $z\sim4$.






\section{Diffuse neutrino expectation from Starburst galaxies}
\begin{figure*} \label{2sigmafig}
  	\centering
  	\includegraphics[width=\linewidth]{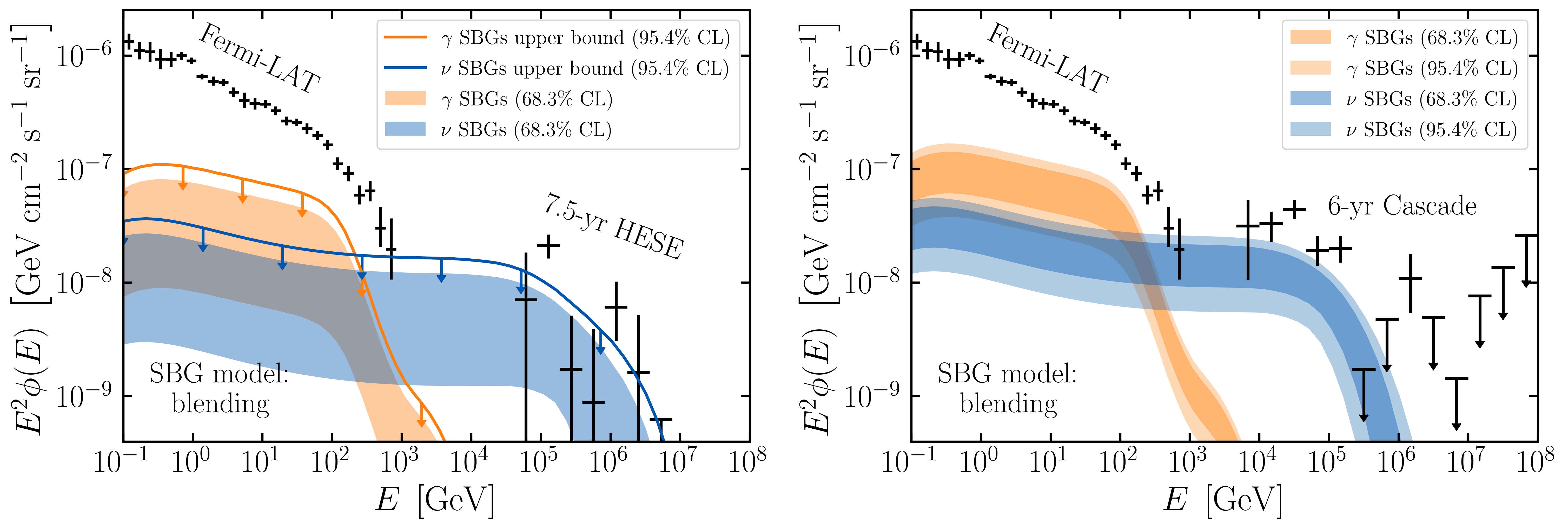}
  	\caption{\label{fig:fluxes_blending_2sigma} Gamma-ray (orange) and single-flavour neutrino (blue) uncertainty bands at $68.3\%$~CL (dark colors) and $95.4\%$~CL (light colors) for the SBG component deduced by the multi-messenger analysis perfomed in Ref. \cite{Ambrosone:2020evo} in case of data-driven blending of spectral indexes. The left (right) plot corresponds to the multi-messenger analysis with IceCube 7.5-year HESE (6-year cascade) neutrino data. In the left plot, the solid lines correspond to upper bounds at $95.4\%$~CL.}
  \end{figure*}

In this section, we outline the theoretical model we used to estimate the Starburst Galaxies neutrino emissions. In particular, we make use of the semi-analytical approach put forward by \cite{Peretti:2019vsj, Ambrosone:2020evo} . We suppose that the star-forming activity for these sources is concentrated on their central core. This particular region is usually called Starburst Nucleus (SBN) and we approximate it to be a spherical region of the order of $250$ pc (see \cite{Peretti:2018tmo} for the details ). As a result, we solve the leaky-box model equation in order to calculate the high-energy proton distribution within the SBN
\begin{equation} \label{leaky}
	F_{p} = Q_{p} \bigg(\frac{1}{T_{adv}} + \frac{1}{T_{loss}} + \frac{1}{T_{diff}}\bigg)^{-1}
\end{equation}
where $F_{p}$ and $Q_{p}$ are respectively the distribution function and the injection rate for protons. Eq. (\ref{leaky}) physically represent the balance between the injection of the high-energy protons given by supernova remnants and the hadronic transport within the SBN itself. In fact, this phenonemon depends on the different timescales: the advection ($T_{adv}$), energy losses ($T_{loss}$) and diffusion processes ($T_{diff}$) (see \cite{Ambrosone:2020evo} for the details).
  Assuming that  protons are  injected  with  a  power-law spectrum in momentum space with spectral index $\alpha$ and an exponential cutoff varying between $1-20 \  \text{PeV}$, the flux is normalized by requiring that each supernova releases into protons $10 \%$ of its total explosion  kinetic  energy  ($10^{51} \ \text{erg}$). We determine the neutrino flux of a single Starburst galaxy using the approach given by \cite{Kelner:2006tc}, which assumes that the pions carry a fixed energy portion of the high-energy protons $(k_{\pi} = 17 \ \%)$.  Assuming the calorimetric condition for the central part of these galaxies, their neutrino flux hardly depends on their structural details and physical parameters; but rather, it is mainly driven by the star formation rate ($\psi$), the cut-off energy ($p_{max}$) of high-energy protons and the spectral index ($\alpha$). Hence, in order to constrain the number of SBGs in the Universe, we use the method of the star formation rate function.  We consider the modified Shechter function $\Phi$SFR(z, $\psi$) reported in \cite{Peretti:2019vsj}, which has
  been obtained by fitting in the redshift interval $0 \le z \le 4.2$ the IR+UV data of a Herschel Source sample after subtracting the AGN contamination. Furthermore, as assumed by \cite{Ambrosone:2020evo}, we do not assume every SBG to have the same spectral index, but instead, we exploit the variability of this parameter along the source class. In particular, we use a Gaussian distribution for these parameters employing the experimental catalogue provided by \cite{Ajello:2020zna}. The formula for the diffuse flux is \cite{Ambrosone:2020evo}
  
  \begin{equation}
  	\begin{aligned}
  		\Phi_{\nu}^\mathrm{SBG} (E, p^\mathrm{max}) = & \int_{0}^{4.2} \mathrm{d}z \int_{\psi_*}^{\infty} \mathrm{d}\log \psi \, \frac{c \, d_c(z)^2}{H(z)} \\ 
  		& \times \Phi_\mathrm{SFR}(z,\psi) \, \Big\langle \phi_{\nu}\big(E,z,\psi,p^\mathrm{max}\big) \Big\rangle_\alpha\,,
  		\label{diffuse}
  	\end{aligned}
  \end{equation}
  where $H(z) = H_0 \sqrt{\Omega_M (1+z)^3 + \Omega_\Lambda}$ is the Hubble parameter with $H_0 = 67.74 \ \mathrm{km} \ \mathrm{s}^{-1}\mathrm{Mpc}^{-1} $, $\Omega_M = 0.31 $ and $\Omega_\Lambda = 0.69$, $\psi_{*} = 2.6 \ \text{M}_{\bigodot} \text{yr}^{-1}$ and $\langle \phi_{\nu} \rangle_\alpha$ is the emitted neutrino fluxes averaged over the distribution of spectral indeces. The astrophysical flux provided by Eq. (\ref{diffuse}) peaks at hudrends of TeVs and it can explain up to $40 \%$ of the IceCube's observations without exceeding the gamma-ray constraints coming from the hadronic component of the extra-galactic background light (EGB).  In fact, in Ref \cite{Ambrosone:2020evo}, a multi-component fit of the IceCube and Fermi-LAT datasets has been perfomed taking into account also Blazars and Radio Galaxies.The main result is highlighted in Fig. \ref{2sigmafig}, which show the $1\sigma$ band and the $2\sigma$ maximum contribution accountable for SBGs; the left panel  corresponds to the multi-messenger analysis with IceCube 7.5-year HESE, while the right one corresponds to the analysis with IceCube 6-year cascade. 
  
As a final remark, we wish to mention that the behaviour of the flux is a more complex function than a simple power-law due to the spectral index blending. However, for energies greater than 100 GeV, it can be accommodated by a simple power-law with an exponential cut-off ($E_{cut}$). In particular, for the forth-coming analysis, we use the expression
\begin{equation}
	\Phi_{\nu} (E) = N \cdot \bigg(\frac{E}{100 \ \text{GeV}}\bigg)^{-2}\cdot E^{-E/E_{cut}}
	\label{flux}
\end{equation}
where $N = 2.74 \cdot 10^{-12}$  GeV$^{-1}$ cm$^{-2}$ s$^{-1}$ and for $E_{cut}$ we use two different values: $0.2 \ $ PeV and $0.5 \ $ PeV which respectively correspond to 5 and 12 PeV for $p_{max}$.  
  
\section{Differential sensitivity with KM3NeT/ARCA telescope}
In order to estimate the probability for ARCA detector to observe the diffuse neutrino signal originated from SBGs was performed the calculation of an appropriate sensitivity. Given the model flux $\Phi_{s}$ by $\ref{flux}$, the sensitivity is calculated at 90 $\%$ of Confidence Level (C.L.) obtained with the Neyman method: 
\begin{equation}
\Phi_{90} = \Phi_{s} \cdot \frac{n_{90}}{n_{s}}
\label{sensitivity}
\end{equation}

\begin{figure*} \label{diffsensy}
  	\centering
  	\includegraphics[width=\linewidth]{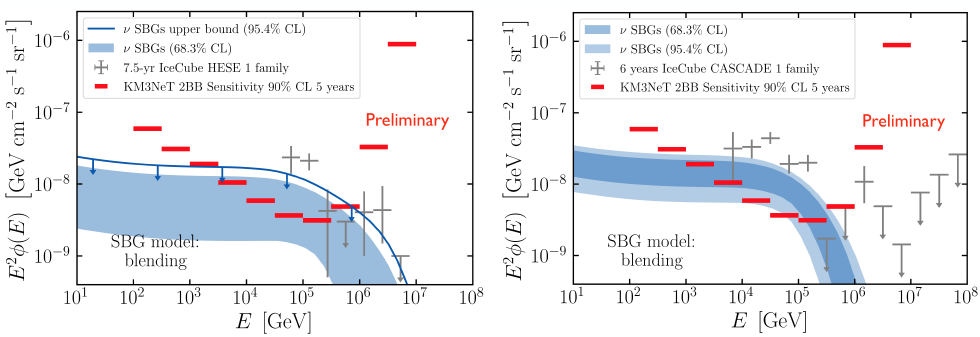}
  	 \caption{In these plots it is reported the differential sensitivity for the flux described by the eq.~\ref{flux} compared with multi-components fitting results from the IceCube HESE sample and the Fermi-LAT EGB, on the left, and  IceCube Cascade sample and  Fermi-LAT EGB.}
  \end{figure*}

The analysis was performed with the latest version of the KM3NeT-ARCA115 MC simulation: the background were defined as atmospheric muons and neutrinos, otherwise the signal was given by the neutrino SBGs (considering for both background and signal only $\nu_{\mu} - \bar{\nu_{\mu}}$ via charge-current interactions). The calculation of the sensitivity involves several assumptions: an interval energy between 100 GeV - 10 PeV was divided in 11 bins in order to calculate a \textit{differential} sensitivity, full sky scenario was considered (no selection in declination), two building blocks of KM3NeT/ARCA detector and five years of acquisition data were take into account. The optimization of this analysis was obtained by applying a bin-per-bin selection chain ($\textbf{Cut\&Count}$ method), considering the following variables: a pre-selection on the reconstructed zenith angle (only events with $\theta_{rec} < 100^{\circ}$ were considered) and a selection on the likelihood ($\lambda$) from track reconstruction algorithm. Finally, once we applied all the selection chain we calculated bin-per-bin the sensitivity reported in eq.~\ref{sensitivity}. \\
The differential KM3NeT/ARCA sensitivity for the described model of SBGs in eq.~\ref{flux} is shown in fig.~\ref{diffsensy}.

\section{Discussion and conclusions}
In this contribution we show how the observations of the incoming KM3NeT/ARCA telescope for the diffuse signal in the range of 100 TeV can be crucial to better constrain the physics of cosmic-ray ``reservoir'' sources. In particular we obtain the expected differential sensitivity of the incoming KM3NeT/ARCA for the diffuse contribution of starburst galaxies up to a redshift of $z\sim 4.2$. Even though differentiate single diffuse astrophysical components can be a challenging purpose, this study shows how few years of data taking of this new telescope can be crucial to select between different multicomponent fitting models.





\section*{The KM3NeT Collaboration author list (July 2021)}
\scriptsize
\noindent
M.~Ageron$^{1}$,
S.~Aiello$^{2}$,
A.~Albert$^{3,55}$,
M.~Alshamsi$^{4}$,
S. Alves Garre$^{5}$,
Z.~Aly$^{1}$,
A. Ambrosone$^{6,7}$,
F.~Ameli$^{8}$,
M.~Andre$^{9}$,
G.~Androulakis$^{10}$
% ...
\\

\noindent
$^{1}$Aix~Marseille~Univ,~CNRS/IN2P3,~CPPM,~Marseille,~France. \\
$^{2}$INFN, Sezione di Catania, Via Santa Sofia 64, Catania, 95123 Italy. \\
$^{3}$Universit{\'e}~de~Strasbourg,~CNRS,~IPHC~UMR~7178,~F-67000~Strasbourg,~France. \\
$^{4}$Universit{\'e} de Paris, CNRS, Astroparticule et Cosmologie, F-75013 Paris, France. \\
$^{5}$IFIC - Instituto de F{\'\i}sica Corpuscular (CSIC - Universitat de Val{\`e}ncia), c/Catedr{\'a}tico Jos{\'e} Beltr{\'a}n, 2, 46980 Paterna, Valencia, Spain. \\
$^{6}$INFN, Sezione di Napoli, Complesso Universitario di Monte S. Angelo, Via Cintia ed. G, Napoli, 80126 Italy. \\
$^{7}$Universit{\`a} di Napoli ``Federico II'', Dip. Scienze Fisiche ``E. Pancini'', Complesso Universitario di Monte S. Angelo, Via Cintia ed. G, Napoli, 80126 Italy. \\
$^{8}$INFN, Sezione di Roma, Piazzale Aldo Moro 2, Roma, 00185 Italy. \\
$^{9}$Universitat Polit{\`e}cnica de Catalunya, Laboratori d'Aplicacions Bioac{\'u}stiques, Centre Tecnol{\`o}gic de Vilanova i la Geltr{\'u}, Avda. Rambla Exposici{\'o}, s/n, Vilanova i la Geltr{\'u}, 08800 Spain. \\
$^{10}$NCSR Demokritos, Institute of Nuclear and Particle Physics, Ag. Paraskevi Attikis, Athens, 15310 Greece. \\
% ...
$^{55}$Universit{\'e} de Haute Alsace, 68100 Mulhouse Cedex, France. \\
% ...

\end{document}